\documentclass[letterpaper]{article}
\usepackage{aaai2026}
\nocopyright
\usepackage{times}
\usepackage{helvet}
\usepackage{courier}
\usepackage[hyphens]{url}
\usepackage{graphicx}
\usepackage{natbib}
\usepackage{caption}
\usepackage{booktabs}
\usepackage{amsmath}
\usepackage{amssymb}
\usepackage{xcolor}
\usepackage{tikz}
\usetikzlibrary{shapes.geometric, arrows.meta, positioning}

\title{AI Watermark Evidence Fails Forensic Readiness: An Empirical Evaluation}
\author{
    Saifur Rahman Tamim, Amir Labib Khan\\
}
\affiliations{
    Department of Computer Science and Engineering\\
    Northern University Bangladesh\\
    tamim\_41230201087@nub.ac.bd, amirlabibkhan02@gmail.com
}

\begin{document}

\maketitle

\begin{center}
\small\itshape
Preprint. A version of this paper was submitted to the AAAI/ACM Conference on AI, Ethics, and Society (AIES) 2026.
\end{center}
\vspace{0.5em}

\begin{abstract}
Governments are increasingly mandating that LLM-generated content carry watermarks. The EU AI Act calls for markings that are ``sufficiently reliable and robust.'' California's SB 942 requires disclosure that is ``permanent or extraordinarily difficult to remove.'' Both mandates rest on an untested assumption: that watermark detection yields evidence reliable enough for courts. This paper tests that assumption directly.

We evaluate three representative LLM watermarking methods---KGW, Unigram, and the MarkLLM implementation of SynthID-Text---against the Daubert admissibility criteria and the NIST SP 800-86 digital forensic process. To structure this evaluation, we propose a Forensic Readiness Score (FRS) framework with 12 criteria, three mandatory gates, and a 60-point scoring system. We focus on meaning-preserving paraphrase as the attack vector, since it is both legally realistic and difficult to dismiss as evidence tampering.

The results raise serious evidentiary concerns. Out of 846 valid paraphrase runs across 15 diverse prompts per method, every single initially-detected KGW and Unigram text lost its watermark after paraphrasing---100\% conditional removal. SynthID fared only slightly better at 98.3\%. Even before any attack, false-negative rates were already high: 70\% for KGW, 83\% for Unigram, 80\% for SynthID. The SynthID configuration also flagged 5.4\% of paraphrased human-written controls as AI-generated and showed an 18.6\% paradox rate, with 80\% of its own pristine watermarked output landing in the uncertainty deadband. None of the three methods satisfy more than two of five Daubert factors. We also find that the FRS point-based scoring system, despite working as designed, cannot fully capture forensic uselessness---a limitation worth noting for future framework design.

These configurations, as tested, do not meet the evidentiary bar that courts require.
\end{abstract}

\section{Introduction}

Consider a courtroom scenario. A prosecutor offers a text as AI-generated, citing a positive watermark detection result. The defense attorney takes that same text, runs it through a paraphrasing tool anyone can access online, and hands back a version that means the same thing---but no longer triggers the watermark detector. The prosecution's evidence just evaporated. The meaning survived. What is the judge supposed to do with that?

This is not a contrived hypothetical. The EU AI Act requires AI-generated content markings to be ``sufficiently reliable, interoperable, effective and robust as far as this is technically feasible'' \cite{euaiact2024}. California's SB 942 insists that disclosure be ``permanent or extraordinarily difficult to remove'' \cite{casb942}. Executive Order 14110 went further, mandating ``state-of-the-art'' provenance tools and explicitly naming watermarking \cite{eo14110}---though the order was later rescinded \cite{eo14179}, which itself says something about how quickly the policy ground can shift.

What these mandates have in common is an assumption: that the underlying watermarking technology actually works well enough to produce courtroom-grade evidence. But this assumption has not been directly evaluated against forensic admissibility standards. Robustness studies measure how well detectors hold up against attacks using ML metrics like TPR and AUC. Governance analyses ask whether the policy infrastructure around watermarking is adequate. Neither tradition provides the thing courts actually need: an end-to-end assessment of whether watermark evidence meets the Daubert admissibility standard and follows the NIST SP 800-86 forensic process \cite{liang2025watermark,nemecek2025watermarking}.

Some background is useful here. The Daubert criteria \cite{daubert1993} are how U.S. federal courts decide whether scientific evidence is admissible---they ask about testability, peer review, known error rates, standards, and general acceptance. NIST SP 800-86 \cite{nist80086} lays out how digital forensic evidence should be collected, examined, analyzed, and reported. Recent work has warned that watermarking risks becoming ``symbolic compliance'' without real enforceable standards \cite{nemecek2025watermarking}. We wanted to see if the empirical data backs up that concern. It does.

This paper contributes four things:
\begin{enumerate}
    \item An empirical evaluation of representative LLM watermarks against forensic admissibility standards (both NIST SP 800-86 and Daubert).
    \item A Forensic Readiness Score (FRS) framework: 12 scorable criteria grounded in Daubert factors and NIST phases, plus 3 mandatory gates and a 60-point scale.
    \item Paraphrase experiments across KGW, Unigram, and SynthID-Text, with reproducibility verified through timestamped same-seed computational reruns.
    \item Evidence that meaning-preserving paraphrase achieves 100\% conditional watermark removal for two of three methods (98.3\% for the third), while keeping semantic similarity high enough that a court would struggle to call it evidence tampering.
\end{enumerate}

As watermarked AI systems reach wider deployment, courts will encounter this kind of evidence without any established framework for evaluating it. We provide that framework---and show that, at least for these configurations, the evidence does not hold up.

\section{Background and Related Work}

\subsection{LLM Watermarking Methods}

We test three watermarking methods that represent different design philosophies, all available through the open-source MarkLLM toolkit \cite{pan2024markllm}.

\textbf{KGW} \cite{kirchenbauer2023watermark} works by splitting the token vocabulary into ``green'' and ``red'' lists using a hash of the previous token. During text generation, green-list tokens get a logit boost controlled by a parameter $\delta$, pushing the model toward those tokens. Detection then checks whether the output has an unusually high proportion of green tokens via a $z$-score. One important consequence of this design: because the partition depends on each token's predecessor, changing even one token can ripple through the rest of the partitions.

\textbf{Unigram} \cite{zhao2024provable} takes a different approach---it uses a fixed, context-independent vocabulary partition. The green/red split stays the same regardless of what came before, which in theory makes it more resilient to local edits. Detection still uses a $z$-score over green-token frequency, but the fixed partition means the method should be more robust to bounded edit-distance attacks.

\textbf{SynthID-Text} \cite{dathathri2024synthid} is the highest-profile method in our evaluation---published in \emph{Nature}, deployed by Google in production. It uses tournament-based scoring with probabilistic detection. We evaluate the open-source MarkLLM implementation, not Google's proprietary system (an important distinction we return to in the limitations). Unlike the other two methods, SynthID uses three detection states---WATERMARKED, NOT\_WATERMARKED, and UNCERTAIN---rather than a simple binary threshold. In our experiments we use the MarkLLM weighted-mean detector with a threshold of 0.5 and an UNCERTAIN deadband of $\pm 0.03$.

These three methods together cover the main design space: context-dependent partitioning (KGW), context-free partitioning (Unigram), and tournament-based probabilistic scoring (SynthID). That range matters because we want to test whether forensic failure is a property of the general approach, not just one particular design.

\subsection{Forensic Evidence Standards}

Two frameworks are relevant to how courts handle scientific and digital evidence.

\textbf{Daubert criteria.} Under \emph{Daubert v. Merrell Dow Pharmaceuticals} \cite{daubert1993} and Federal Rule of Evidence 702 \cite{fre702}, judges evaluate scientific evidence on five factors: (1) testability and falsifiability, (2) peer review and publication, (3) known or potential error rate, (4) existence of controlling standards, and (5) general acceptance. For watermarking, Factor 3---the known error rate---turns out to be the critical weakness. If the error rate is unstable or unknown, the evidence is hard to admit. We note that Daubert is a US federal evidentiary standard; state courts following \emph{Frye} and civil-law jurisdictions (e.g., under EU procedural rules) use different admissibility tests. We use Daubert as our primary lens because of its close conceptual fit with NIST's error-rate language, but the forensic weaknesses we document---unstable error rates, no controlling standards---are relevant to any admissibility framework that asks similar questions of scientific evidence.

\textbf{NIST SP 800-86.} NIST's guide for digital forensics \cite{nist80086} defines four phases: Collection, Examination, Analysis, and Reporting. None of the watermark evaluations we are aware of document an end-to-end workflow aligned to these phases. This matters because watermark detection alone is not a forensic process; it becomes forensic evidence only when embedded in a documented collection, examination, analysis, and reporting workflow. NIST has separately acknowledged the broader challenge of digital content transparency \cite{nistai1004}, but no existing guidance bridges watermark detection to forensic evidence handling.

\textbf{Historical precedent.} There is a pattern here worth noting. The 2009 NAS report \cite{nas2009} found that several forensic disciplines---fingerprint analysis, bite mark comparison, hair microscopy---had been used in courtrooms for decades without adequate scientific validation. The 2016 PCAST report \cite{pcast2016} made similar points. The consequences were real: wrongful convictions. AI watermarking looks like it could be heading down the same path---deployed in practice, written into regulation, but not actually validated against the standards courts use to decide whether evidence is reliable enough to hear.

\subsection{Related Work}

\textbf{Watermark robustness evaluation.} The most thorough robustness study to date is WaterPark by \citet{liang2025watermark}, which tested 10 watermarking methods against 12 attack types across three language models (OPT-1.3B, LLaMA3-7B, Qwen2.5-14B) and five datasets. Their findings are directly relevant to ours: SynthID's true positive rate dropped from 0.998 on clean text to 0.498 under moderate paraphrasing. A single ChatGPT paraphrase pass brought \emph{every} tested method below 30\% detection. They also found enormous cross-model variability---KGW's TPR went from 0.858 on OPT to 0.334 on LLaMA3 under the same attack conditions. But WaterPark evaluates entirely through ML metrics. It does not ask whether these numbers would satisfy a Daubert hearing, and it does not map results to NIST forensic phases. Other work has established theoretical impossibility results for strong watermarking \cite{zhang2024watermarks}, shown that reliable AI text detection may be impossible in adversarial settings \cite{sadasivan2023detecting}, and demonstrated attacks that exploit watermark design properties \cite{pang2024attacking}. These all point toward fragility, but none address admissibility. Notably, the WaterPark finding that one round of ChatGPT paraphrasing drops all methods below 30\% TPR suggests our same-model paraphraser actually represents a \emph{conservative} lower bound on how bad things can get.

\textbf{Distillation-based attacks and attribution ambiguity.} \citet{pan2025watermarks} demonstrated that watermark traces can be stripped away during unauthorized knowledge distillation. \citet{yi2025unified} extended this to unified spoofing and scrubbing attacks across KGW, Unigram, and SynthID-Text. The forensic implication goes beyond robustness: even when a watermark \emph{is} detected, the signal might not mean what you think it means. It could reflect direct generation, inherited traces from distillation, or deliberate spoofing. That kind of attribution ambiguity makes it very difficult to establish a stable error-rate interpretation in court.

\textbf{Governance and policy.} \citet{nemecek2025watermarking} argued that watermarking without enforceable standards amounts to ``symbolic compliance'' and laid out a three-layer governance framework covering technical requirements, audit infrastructure, and enforcement. Their paper included a prototype evaluation scorecard with 0--5 scoring---a design that ended up converging with our FRS. They also ran SynthID on Gemma-2-9b-it independently and found paraphrasing eliminated detection in 4 of 5 cases. But their contribution is fundamentally a governance argument, not a systematic empirical test: five prompts, no forensic pipeline, no FRS-style scoring.

\begin{table}[t]
\centering
\small
\caption{How this paper fits alongside prior watermark evaluations. We go narrower on attacks but deeper on forensic admissibility---to our knowledge, the first study to jointly apply Daubert and NIST SP 800-86 forensic standards to LLM watermark evaluation, building on prior robustness (Liang et al.) and governance (Nemecek et al.) work.}
\label{tab:comparison}
\resizebox{\columnwidth}{!}{%
\begin{tabular}{@{}llll@{}}
\toprule
\textbf{Work} & \textbf{Scale} & \textbf{Attacks} & \textbf{Forensic framework} \\
\midrule
Liang et al.\            & 10 methods & 12 types     & ML metrics (TPR, FPR) \\
Nemecek et al.\          & 1 method   & 1 type       & Governance scorecard \\
Pan et al.; Yi et al.\   & 3 methods  & Distillation & Not addressed \\
\textbf{This work}       & 3 methods  & 1 type       & FRS + Daubert + NIST \\
\bottomrule
\end{tabular}%
}
\end{table}

\textbf{Our position.} There are three threads in the existing literature. \citet{liang2025watermark} ask: how robust are watermarks? \citet{nemecek2025watermarking} ask: why is governance failing? Distillation-attack studies ask whether watermark traces survive model transfer \cite{pan2025watermarks,yi2025unified}. We ask a different question: \emph{would watermark evidence actually survive a Daubert hearing?} For the three methods and configurations we tested, the answer is no.

\section{The Forensic Readiness Score Framework}

ML benchmarks ask: ``does watermark detection work on average?'' Courts need to know something different: ``can an expert witness testify, for \emph{this specific} piece of evidence, to a known and stable error rate?'' The FRS framework is our attempt to bridge that gap.

\subsection{Design Rationale}

The FRS maps NIST SP 800-86's four phases onto five framework components (Evidence Collection, Verification Protocol, FRS Evaluation, Evidence Report Template, and Implementation Guidelines) and turns Daubert's five factors into scorable criteria. We were not the first to think along these lines---\citet{nemecek2025watermarking} independently proposed a scorecard with 0--5 scoring across robustness, detection quality, and auditability. Our FRS shares that philosophy but takes it further: each criterion is grounded in a specific Daubert factor or NIST phase, we add mandatory gates that can override point scores entirely, and we validate the whole thing empirically.

\subsection{The 12 Criteria}

Each criterion gets a score from 0 to 5, for a maximum of 60 points. They fall into three categories (Table~\ref{tab:criteria}).

\begin{table}[t]
\centering
\small
\caption{FRS criteria mapped to forensic standards.}
\label{tab:criteria}
\begin{tabular}{@{}lp{0.4cm}p{2.5cm}p{2.0cm}@{}}
\toprule
\textbf{Cat.} & \textbf{\#} & \textbf{Criterion} & \textbf{Standard} \\
\midrule
Tech. & T1 & Repeatability & NIST reprod. \\
 & T2 & Robustness & NIST exam. \\
 & T3 & Detectability & Daubert F3 \\
 & T4 & Quantifiability & Daubert F1 \\
\midrule
Legal & L1 & Reliability & Daubert F3 \\
 & L2 & Peer Review & Daubert F2 \\
 & L3 & Known Error Rate & Daubert F3 \\
 & L4 & Testability & Daubert F1 \\
\midrule
Oper. & O1 & Accessibility & NIST cross. \\
 & O2 & Efficiency & NIST cross. \\
 & O3 & Documentation & NIST cross. \\
 & O4 & Independence & NIST cross. \\
\bottomrule
\end{tabular}
\end{table}

\subsection{The Three Mandatory Gates}

Points alone are not enough. Even if a method scores $\geq$40/60, it still has to clear three gates. Failing any one of them results in a NOT FORENSIC READY verdict, no matter how high the score.

\textbf{Gate G1:} FPR and FNR must be documented and independently computable. Without known error rates, courts cannot evaluate reliability (Daubert Factor 3).

\textbf{Gate G2:} The paradox rate must stay below 20\%. If attacking a watermark \emph{increases} the detection score more than a fifth of the time, something is fundamentally wrong with the method's logic. Evidence corruption should not help verification.

\textbf{Gate G3:} Results must be repeatable across sessions within documented tolerance. This is a basic NIST reproducibility requirement.

\subsection{Scoring Limits as a Finding}

One of the more interesting outcomes of this work is what happens with the scoring system itself. As Section~5 will show, KGW ends up at 37/60---NOT FORENSIC READY, and fairly classified. But Unigram lands at exactly 40/60, right on the CONDITIONALLY FORENSIC READY threshold, despite having the worst FNR of any method (83\%) and 100\% conditional removal. It passes every gate and clears the point threshold while being, in practice, forensically useless.

We think this is worth highlighting not as a flaw in the framework but as a genuine finding: point-based scoring has inherent limits when it comes to capturing forensic uselessness. It is a bit like how a good credit score can mask individual risk. Courts will need both the quantitative score \emph{and} a qualitative assessment of how the method actually behaves under adversarial conditions.

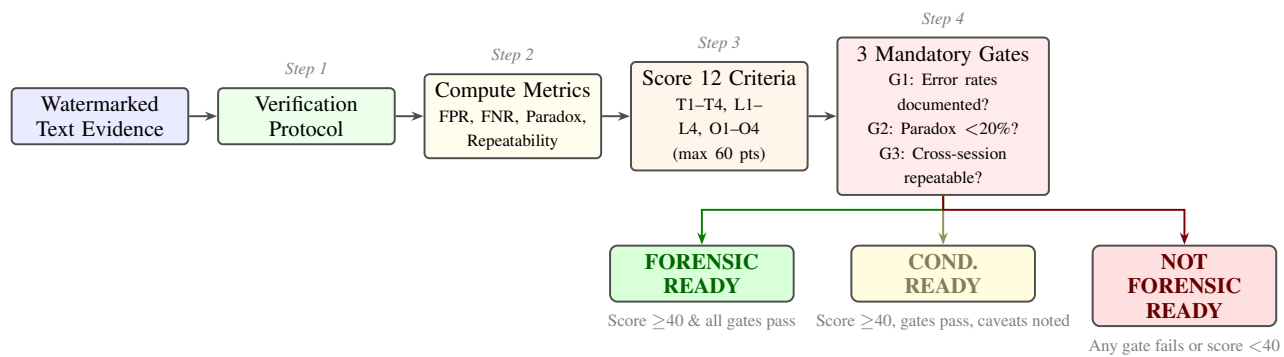
\begin{figure*}[t]
\centering
\resizebox{0.95\textwidth}{!}{%
\begin{tikzpicture}[
  node distance=0.5cm and 0.4cm,
  every node/.style={font=\small},
  process/.style={rectangle, rounded corners=2pt, draw=black!70, thick, 
    minimum height=0.8cm, align=center, text width=2.3cm},
  gate/.style={rectangle, rounded corners=2pt, draw=black!70, thick,
    minimum height=0.8cm, align=center, text width=2.8cm},
  verdict/.style={rectangle, rounded corners=3pt, draw=black!70, thick,
    minimum height=0.65cm, align=center, text width=2.4cm, font=\small\bfseries},
  arr/.style={-{Stealth[length=5pt]}, thick, black!70},
  label/.style={font=\scriptsize, text=black!50}
]

\node[process, fill=blue!8] (input) {Watermarked\\Text Evidence};

\node[process, fill=green!8, right=of input] (verify) {Verification\\Protocol};

\node[process, fill=yellow!8, right=of verify] (metrics) {Compute Metrics\\{\scriptsize FPR, FNR, Paradox,}\\{\scriptsize Repeatability}};

\node[process, fill=orange!8, right=of metrics] (score) {Score 12 Criteria\\{\scriptsize T1--T4, L1--L4, O1--O4}\\{\scriptsize (max 60 pts)}};

\node[gate, fill=red!8, right=of score] (gates) {3 Mandatory Gates\\{\scriptsize G1: Error rates documented?}\\{\scriptsize G2: Paradox $<$20\%?}\\{\scriptsize G3: Cross-session repeatable?}};

\draw[arr] (input) -- (verify);
\draw[arr] (verify) -- (metrics);
\draw[arr] (metrics) -- (score);
\draw[arr] (score) -- (gates);

\node[verdict, fill=green!15, below left=0.7cm and 0.6cm of gates] (ready) 
  {\color{green!40!black}FORENSIC\\READY};
\node[label, below=0.05cm of ready] (r_desc) {Score $\geq$40 \& all gates pass};

\node[verdict, fill=yellow!15, below=0.7cm of gates] (cond) 
  {\color{yellow!40!black}COND.\\READY};
\node[label, below=0.05cm of cond] (c_desc) {Score $\geq$40, gates pass, caveats noted};

\node[verdict, fill=red!12, below right=0.7cm and 0.6cm of gates] (notready) 
  {\color{red!40!black}NOT FORENSIC\\READY};
\node[label, below=0.05cm of notready] (n_desc) {Any gate fails or score $<$40};

\draw[arr, green!50!black] (gates.south) -- ++(0,-0.2) -| (ready.north);
\draw[arr, yellow!50!black] (gates.south) -- (cond.north);
\draw[arr, red!50!black] (gates.south) -- ++(0,-0.2) -| (notready.north);

\node[label, above=0.01cm of verify.north] {\textit{Step 1}};
\node[label, above=0.01cm of metrics.north] {\textit{Step 2}};
\node[label, above=0.01cm of score.north] {\textit{Step 3}};
\node[label, above=0.01cm of gates.north] {\textit{Step 4}};

\end{tikzpicture}%
}
\caption{How the FRS evaluation works. Watermarked text moves through four stages: verification testing, metric computation, scoring against 12 criteria (0--5 each, 60 max), and three mandatory gates. Any single gate failure overrides the point score---a method that fails a gate is NOT FORENSIC READY no matter how many points it earned. In our results, Unigram passes all gates and lands at exactly 40/60 (the minimum COND. READY threshold) despite 100\% conditional removal and 83\% FNR---a clear scoring-limit case.}
\label{fig:frs_pipeline}
\end{figure*}

\section{Experimental Setup}

\subsection{Methods and Models}

We run all experiments through the MarkLLM toolkit \cite{pan2024markllm}. Table~\ref{tab:setup} summarizes the configuration.

\begin{table}[t]
\centering
\caption{Experimental configuration. A single attack model (Qwen2.5-1.5B) is used across all methods, controlling the attacker variable so that cross-method differences reflect watermark properties rather than paraphraser capability.}
\label{tab:setup}
\resizebox{\columnwidth}{!}{%
\begin{tabular}{@{}lllll@{}}
\toprule
\textbf{Method} & \textbf{Source Model} & \textbf{Attack Model} & \textbf{Venue} & \textbf{Detector} \\
\midrule
KGW & Qwen2.5-1.5B & Qwen2.5-1.5B & ICML '23 & $z$-score ($\tau$=4.0) \\
Unigram & Qwen2.5-1.5B & Qwen2.5-1.5B & ICLR '24 & $z$-score ($\tau$=4.0) \\
SynthID & Gemma-2-9b-it & Qwen2.5-1.5B & Nature '24 & weighted\_mean ($\tau$=0.5$\pm$0.03) \\
\bottomrule
\end{tabular}%
}
\end{table}

For KGW and Unigram, Qwen2.5-1.5B serves as both the watermark generator and the paraphrase attacker, which keeps the comparison clean across methods. For SynthID, we use Gemma-2-9b-it as the generator---following the setup in \citet{nemecek2025watermarking}---and Qwen2.5-1.5B as a cross-model attacker. This is actually a more realistic threat model: in practice, someone trying to strip a watermark does not need access to the original model. The fact that we use different source models for different methods is deliberate. We are testing whether forensic failure is a property of statistical watermarking \emph{as a category}, not something specific to one model.

\subsection{Attack Design}

\textbf{Meaning-preserving paraphrase.} Qwen2.5-1.5B rewrites each watermarked text for all three methods. For SynthID, this creates a cross-model attack (Gemma-generated text attacked by Qwen); for KGW and Unigram, it is a same-model attack. We filter outputs through a triple validity gate: (1) cosine similarity $\geq 0.75$ via all-MiniLM-L6-v2 \cite{reimers2019sentence}, (2) normalized Levenshtein distance $\geq 0.15$, and (3) length ratio between 0.5 and 2.0. Anything failing any gate gets thrown out.

The retained paraphrases ended up with high semantic similarity---medians of 0.83--0.84 across methods, with every single retained sample above 0.75. We chose paraphrasing as the sole attack because it is the one that matters most in a legal setting. A defense attorney can point to a paraphrased text and say: ``the meaning is identical---you cannot call this evidence destruction.'' A court would have a very hard time excluding that argument.

\subsection{Experimental Protocol}

We designed 15 prompts per method, covering conversational, technical, news, professional, and creative domains. Each prompt gets 2 generation seeds, giving 30 base watermarked texts per method. We then attack each base text at three temperatures (0.7, 1.0, 1.3) with five template variants per temperature, which means up to 450 attack attempts per method before filtering. After the validity gate, we end up with 304 valid runs for KGW, 306 for Unigram, and 236 for SynthID. We use prompt-level statistics as the primary unit of analysis and compute SHA-256 hashes at every stage.

For baseline error rates, the 30 pristine texts per method give us FNR estimates. Paraphrased controls---237 for KGW and Unigram, 184 for SynthID---provide FPR baselines. The Qwen experiments ran on Colab free tier (T4 GPU); SynthID needed Colab Pro for the A100 to handle Gemma-2-9b-it.

\begin{table}[t]
\centering
\small
\caption{Session provenance. Replicate sessions used identical seeds and produced byte-identical results, confirming deterministic reproducibility. We report canonical session results; replicate sessions are archived for provenance.}
\label{tab:provenance}
\begin{tabular}{@{}llll@{}}
\toprule
\textbf{Session} & \textbf{Notebook} & \textbf{Seed} & \textbf{Status} \\
\midrule
qwen\_v4.1\_rep2 & KGW+Unigram & 123 & Canonical \\
qwen\_v4.1\_rep3 & KGW+Unigram & 123 & Reproducibility \\
synthid\_v4 & SynthID & 123 & Canonical \\
synthid\_v4\_rep2 & SynthID & 123 & Reproducibility \\
\bottomrule
\end{tabular}
\end{table}

Every result we report comes from archived, version-controlled runs verified with SHA-256 hashes (Table~\ref{tab:provenance}). We ran each experiment twice under identical seeds and configuration; the reruns produced byte-identical outputs and serve as deterministic reproducibility evidence, not additional independent samples. Code and archived artifacts will be released upon acceptance.

\section{Results}

The results tell a story in three layers: watermarks fail before any attack, they collapse under paraphrase, and the paraphrases preserve enough meaning that a court cannot dismiss them as evidence destruction.

\subsection{Pre-Attack Baseline Failure}

Before anyone even tries to attack the watermarks, they are already unreliable. The fraction of watermarked texts that the detectors correctly identify---with no adversarial intervention at all---is surprisingly low. KGW picks up 9 out of 30 pristine texts (FNR = 70\%, Wilson 95\% CI: 52.1--83.3\%). Unigram does worse: 5 out of 30 (FNR = 83\%, CI: 66.4--92.7\%). SynthID manages 6 out of 30 (FNR = 80\%, CI: 62.7--90.5\%).

For SynthID the picture is particularly bleak: 24 of those 30 pristine texts scored inside the $\pm 0.03$ uncertainty deadband. The detector fails to issue a confident verdict on 80\% of its own watermarked output.

This matters for Daubert. If a detector cannot reliably find its own watermarks in clean, unmodified text, an expert witness cannot credibly testify to a ``known error rate.'' The evidence is already compromised before any adversary gets involved.

\subsection{Paraphrase Attack: Primary Analysis}

Table~\ref{tab:paraphrase} has the core numbers. Among texts where the watermark was initially detected, paraphrasing wiped it out in every single case for KGW and Unigram. For SynthID, 58 out of 59 initially-detected texts lost their watermark---98.3\%.

\begin{table}[t]
\centering
\caption{Paraphrase attack results across three methods. Conditional removal is the headline finding: among initially detectable texts, what fraction lost detection after paraphrase. Wilson 95\% CIs reported for FPR, FNR, and conditional removal.}
\label{tab:paraphrase}
\resizebox{\columnwidth}{!}{%
\begin{tabular}{@{}lccc@{}}
\toprule
\textbf{Metric} & \textbf{KGW} & \textbf{Unigram} & \textbf{SynthID} \\
\midrule
Valid runs & 304 & 306 & 236 \\
Yield & 34\% & 34.5\% & 23\% \\
Raw removal & 32.9\% & 13.4\% & 24.6\% \\
\textbf{Cond.\ removal} & \textbf{100\%} (100/100) & \textbf{100\%} (41/41) & \textbf{98.3\%} (58/59) \\
FPR (controls) & 0/237 (0\%) & 0/237 (0\%) & \textbf{10/184 (5.4\%)} \\
FPR 95\% CI & [0, 1.6]\% & [0, 1.6]\% & [3.0, 9.7]\% \\
FNR (pristine) & 21/30 (70\%) & 25/30 (83\%) & 24/30 (80\%) \\
FNR 95\% CI & [52.1, 83.3]\% & [66.4, 92.7]\% & [62.7, 90.5]\% \\
Paradox rate & 2.3\% & 4.9\% & 18.6\% \\
Sem.\ sim.\ (median) & 0.84 & 0.83 & 0.83 \\
\bottomrule
\end{tabular}%
}
\end{table}

\begin{figure}[t]
\centering
\includegraphics[width=\linewidth]{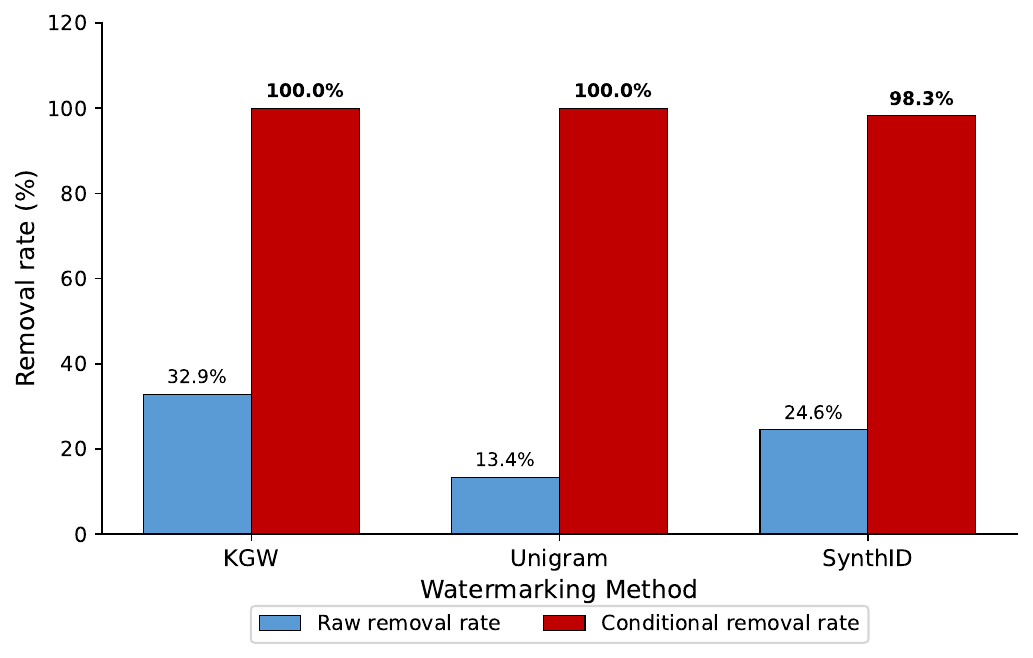}
\caption{Raw vs.\ conditional watermark removal rates. Raw removal (blue) counts all texts losing detection; conditional removal (red) counts only texts that were initially detectable. The gap exposes the forensic problem: raw rates appear modest (13--33\%), but among the small fraction of texts a court could actually rely on, removal is near-total (98--100\%).}
\label{fig:raw_vs_cond}
\end{figure}

That 100\% conditional removal for KGW and Unigram is the central result of this paper (Figure~\ref{fig:raw_vs_cond}). What it means concretely: for any watermarked text that a forensic examiner could actually detect, a straightforward paraphrase will erase the watermark entirely. The paraphrasing tools needed are freely available. The meaning is preserved.

We should note that we only discarded paraphrases when they failed the validity gate (similarity below 0.75 or too little lexical change), so Table~\ref{tab:paraphrase} represents the conservative, high-quality subset that would be hardest for a prosecutor to challenge.

\textbf{The compounding problem.} These failures stack. KGW detects 30\% of its own watermarked texts to begin with, and then loses all of those to paraphrase: $0.30 \times 0.00 = 0.00$ effective detection. Unigram is even worse---17\% baseline detection, then 0\% survival. SynthID detects 20\% and retains 1.7\% of those: $0.20 \times 0.017 \approx 0.003$. A court relying on watermark evidence is looking at effective detection rates somewhere between 0\% and 0.3\% after a trivial, meaning-preserving transformation.

\textbf{SynthID-specific findings.} SynthID has five forensically noteworthy behaviors.

First, \emph{phantom watermarks}: the detector flagged 10 out of 184 paraphrased control texts---clean, unwatermarked human text---as watermarked (5.4\%, Wilson CI: 3.0--9.7\%). That means human-authored text can be falsely attributed to an AI system. One of the flagged samples was a standard typing test sentence, which scored 0.5656 and cleared the detection threshold. If a standard typing test sentence can exceed the detection threshold, the false-attribution risk is non-trivial.

Second, 24 out of 30 pristine watermarked texts (80\%) scored inside the $\pm 0.03$ uncertainty deadband. The detector's usable operating range is about 0.07 points wide (0.487 to 0.558). There is almost no room between ``watermarked'' and ``not watermarked.''

Third, \emph{threshold sensitivity}: if you shift the threshold by just $\pm$10\%, 93.6\% of verdicts flip. That earns a T4 (quantifiability) score of 0/5 in our FRS. There is no stable quantitative signal to testify about.

Fourth, after paraphrasing, 69.5\% of attacked texts stay stuck in the UNCERTAIN zone while 23.7\% transition from DETECTED to UNCERTAIN. The three-state system, which was supposed to reduce false certainty, instead creates a forensic dead zone.

Fifth, despite being published in \emph{Nature} and deployed by Google, the MarkLLM SynthID configuration scores lowest of all three methods at 35/60. \citet{liang2025watermark} independently found SynthID's TPR dropping to 0.498 under DP-40 paraphrase and 0.232 under translation, confirming that the method works on margins too thin for forensic use.

\subsection{Semantic Preservation}

The retained paraphrases stay semantically close to the originals: median cosine similarities of 0.84 (KGW), 0.83 (Unigram), and 0.83 (SynthID), with everything above 0.75. This is what makes the attack so damaging from a legal perspective. You cannot call something ``evidence destruction'' when the meaning is substantially intact. The transformation is just rewording---the kind of thing people do naturally every day.

\begin{figure*}[t]
\centering
\includegraphics[width=0.95\textwidth]{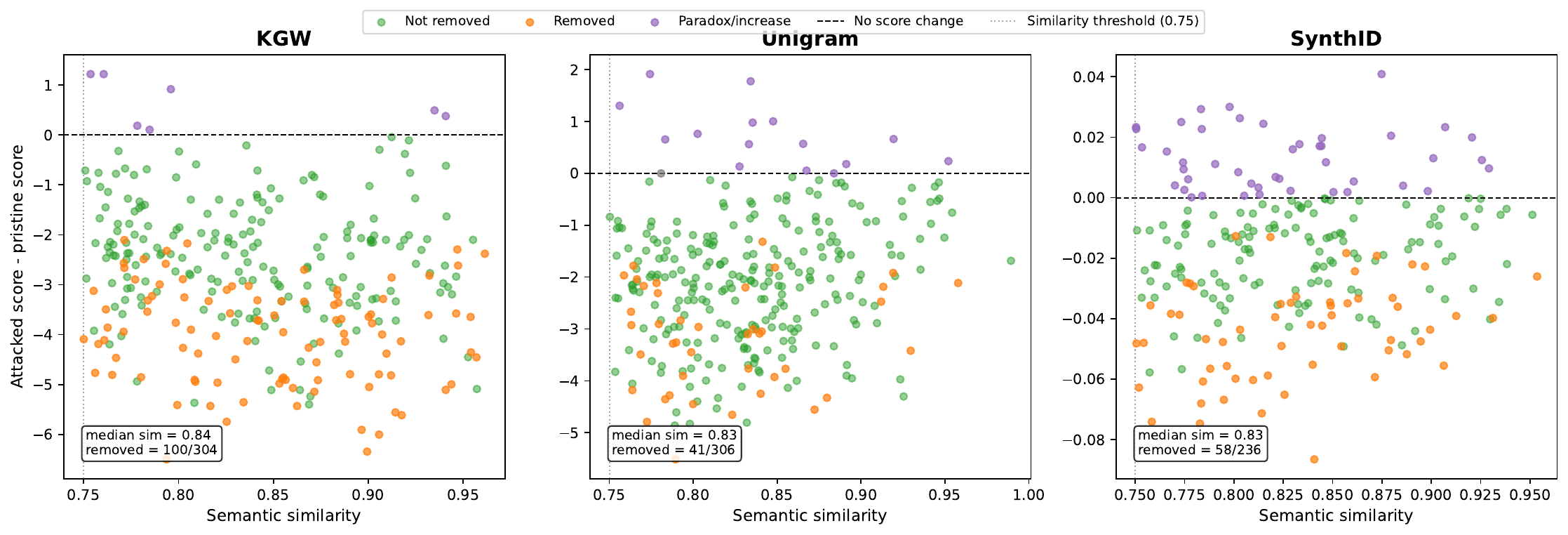}
\caption{Within-method score change under paraphrase attack for KGW, Unigram, and SynthID. Each point is a valid paraphrase run; the x-axis gives semantic similarity and the y-axis gives attacked score minus pristine score. Orange points denote removal events, green points denote non-removal, and purple points denote paradoxical score increases. Across all three methods, score drops persist even when semantic similarity remains high ($\geq 0.75$), supporting the forensic claim that watermark disruption does not require meaning destruction. Score scales differ by method, so comparisons are interpreted within each panel rather than across panels. $n$ = 304 (KGW), 306 (Unigram), 236 (SynthID).}
\label{fig:paraphrase_diagnostics}
\end{figure*}

\begin{figure}[t]
\centering
\includegraphics[width=\linewidth]{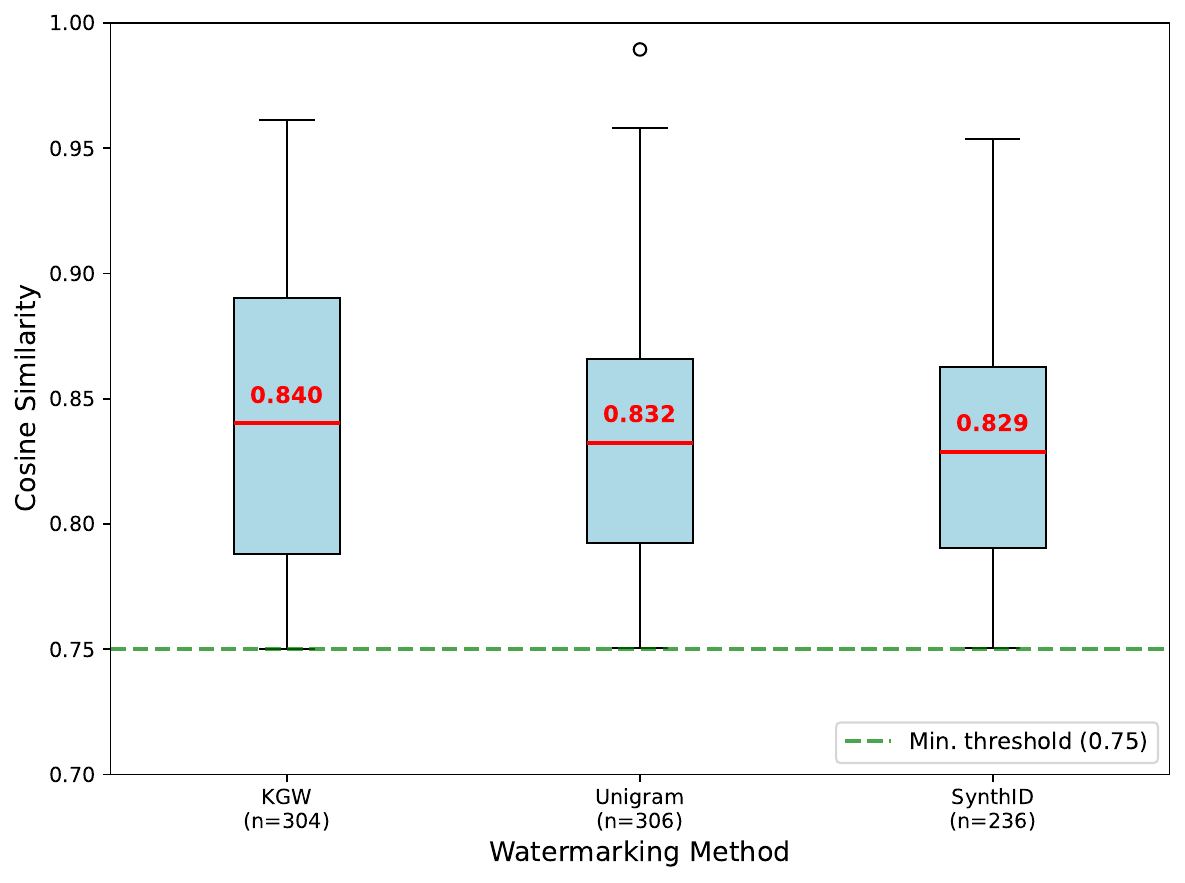}
\caption{Semantic similarity distribution across paraphrase attacks (all three methods). Box plots show median, quartiles, and range of cosine similarity for retained paraphrases. The green dashed line indicates the minimum acceptable semantic threshold (0.75). All methods show medians between 0.83--0.84 and retain samples exclusively above 0.75, confirming that watermark removal occurs without meaning destruction. $n$ = 304 (KGW), 306 (Unigram), 236 (SynthID).}
\label{fig:semantic_distribution}
\end{figure}

We also ran preliminary word deletion experiments on earlier pilot datasets and saw the same general failure patterns, but we chose to focus the paper entirely on paraphrase since it is the legally realistic scenario. Our results line up with \citet{liang2025watermark}: their WaterPark evaluation tested 12 attack types across three LLMs and found fragility everywhere. Their SynthID TPR dropped to 0.498 under moderate paraphrasing, 0.232 under translation. And a single ChatGPT paraphrase pass brought every method below 30\% TPR---meaning our same-model attacker is actually a conservative test.

\subsection{FRS Audit and Daubert Scorecard}

Table~\ref{tab:frs} gives the FRS scores as an interpretive lens over the empirical results. The experiments in Sections 5.1--5.3 are the primary evidence; FRS adds structure.

\begin{table}[t]
\centering
\caption{FRS audit results under paraphrase attack. The ``Empirical reality'' column contextualizes the FRS score---methods that pass the point threshold can still be forensically useless.}
\label{tab:frs}
\resizebox{\columnwidth}{!}{%
\begin{tabular}{@{}lcccl@{}}
\toprule
\textbf{Method} & \textbf{FRS} & \textbf{Gates} & \textbf{Verdict} & \textbf{Empirical reality} \\
\midrule
KGW & 37/60 & All pass & Not ready & 100\% cond.\ removal, 70\% FNR \\
Unigram & 40/60 & All pass & Cond.\ ready & 100\% cond.\ removal, 83\% FNR \\
SynthID & 35/60 & All pass & Not ready & 5.4\% FPR, 80\% UNCERTAIN, T4=0 \\
\bottomrule
\end{tabular}%
}
\end{table}

KGW at 37/60 is correctly flagged as NOT FORENSIC READY. So is SynthID at 35/60---it fails on score alone, and the empirical picture (false positives, near-total uncertainty) is even worse than the number suggests.

The interesting case is Unigram. It lands at exactly 40/60---the minimum passing score---and clears all three gates. By the FRS rubric, it is CONDITIONALLY FORENSIC READY. But it has the highest FNR (83\%), 100\% conditional removal, and the worst baseline detection of any method. One point less and it would be NOT READY. This is the clearest demonstration we have that point-based forensic scoring can mask underlying uselessness.

Table~\ref{tab:daubert} maps each method against the five Daubert factors. All three pass Factors 1 and 2 (they are testable and peer-reviewed). All three fail Factor 3 (no stable error rate) and Factor 4 (no forensic standards existed before this work). Factor 5 gets a ``partial''---accepted in the research community but not in forensic practice. To be clear: we do not mean that error rates cannot be measured in a controlled experimental notebook. Rather, the measured rates are high, attack-sensitive, and configuration-dependent, preventing a stable operational error rate from being generalized to case-level forensic use under Rule 702. Passing only two out of five Daubert factors makes admissibility a steep climb.

\begin{table}[t]
\centering
\small
\caption{Daubert factor assessment by method. ``Part.'' denotes partial satisfaction: accepted in research literature, but not in forensic practice.}
\label{tab:daubert}
\begin{tabular}{@{}lccccc@{}}
\toprule
\textbf{Method} & \textbf{F1} & \textbf{F2} & \textbf{F3} & \textbf{F4} & \textbf{F5} \\
\midrule
KGW & Pass & Pass & Fail & Fail & Part. \\
Unigram & Pass & Pass & Fail & Fail & Part. \\
SynthID & Pass & Pass & Fail & Fail & Part. \\
\bottomrule
\end{tabular}
\end{table}

\section{Discussion}

\subsection{Implications for Courts}

We want to be precise about what these experiments show. They do not show that watermark detectors are stochastically unreliable in the sense of producing random outputs. The detectors are deterministically repeatable---run the same input twice, get the same score. What collapses is the \emph{evidentiary signal}: once a text undergoes meaning-preserving transformation, the detection result changes. This is reproducible forensic failure, not noise.

Judges are going to encounter watermark evidence. The FRS framework gives them a structured way to evaluate it using the forensic language they already work with. But the bottom line is simpler than the framework: until watermarking methods can demonstrate stable error rates under realistic adversarial conditions, watermark evidence deserves serious skepticism.

There is a broader issue, too. Even a positive watermark signal may not settle the authorship question. Distillation and spoofing research shows that a detected watermark might reflect direct generation, inherited traces from model distillation, or deliberate forgery \cite{pan2025watermarks,yi2025unified}. Watermarking sits inside a larger, unsolved problem of digital evidence authentication in the AI era \cite{bellovin2024seeking}.

Courts have been down this road before. The NAS report \cite{nas2009} documented how forensic methods were admitted and relied upon for years before anyone checked whether the science held up. We think AI watermarking is at risk of repeating that mistake.

\subsection{Implications for Policymakers}

The EU AI Act \cite{euaiact2024} and California's SB 942 \cite{casb942} both mandate watermarking, but every method we tested fails forensic admissibility standards in our evaluated configurations. This is essentially what \citet{nemecek2025watermarking} called ``symbolic compliance''---the mandate exists, implementations exist, but the connection between the two has not been validated. \citet{liang2025watermark} showed that watermark designers face a basic trade-off: methods that keep text quality high (like KGW and Unigram) tend to be fragile, while robust methods degrade quality. And the cross-model variability they documented (TPR ranging from 0.858 to 0.334 for the same method on different models) makes it essentially impossible to establish a stable ``known error rate'' across deployments---a direct Daubert Factor 3 violation.

\subsection{Implications for AI Companies}

The MarkLLM SynthID configuration scored lowest of the three methods (35/60), with a detection margin of roughly 0.03, a 5.4\% false positive rate, 80\% UNCERTAIN verdicts on its own pristine output, and T4 = 0 (93.6\% verdict instability under threshold perturbation). A standard human typing test sentence scored above the detection threshold. The pattern of findings across our work, \citet{nemecek2025watermarking}, and \citet{liang2025watermark} all point in the same direction: deployment interest has gotten ahead of forensic validation.

\subsection{The Scoring Limit Finding}

Unigram's exact-threshold pass (40/60) while exhibiting the worst empirical performance of any method---100\% conditional removal, 83\% FNR---is a finding in its own right. It shows that criteria-based scoring frameworks, no matter how carefully designed, have inherent limits. Forensic validation for AI systems needs adversarial testing \emph{alongside} the scorecard, not folded into it.

\section{Limitations and Future Work}

\textbf{Model scale and families.} We tested two model families (Qwen2.5 and Gemma-2) at 1.5B--9B scale. \citet{liang2025watermark} found cross-model fragility persists at 14B, and the statistical properties underlying these failures are scale-independent. Testing additional families like LLaMA or Mistral would widen coverage, but given the category-wide nature of the failure, we would not expect different forensic conclusions.

\textbf{Attack scope.} We tested one attack type. \citet{liang2025watermark} tested twelve and found consistent fragility everywhere. Their ChatGPT paraphrase result---all methods dropping below 30\% TPR---suggests stronger attackers would produce worse outcomes than what we report. We also ran earlier word deletion pilot experiments that informed our anomaly taxonomy but used a different experimental protocol, so we do not report those as primary results.

\textbf{Text length and sample size.} Our texts are 50 tokens long, and we use 30 pristine texts (15 prompts $\times$ 2 seeds) per method. Longer texts and larger sample sets are needed in future work.

\textbf{Reproducibility scope.} All experiments use a single global seed (123). Our replicate sessions confirm deterministic reproducibility, but they are not cross-seed replications. We do have cross-seed data from earlier pilot experiments (seed 42 vs.\ 123 for KGW/Unigram), which showed identical failure patterns, but this was under a slightly different protocol.

\textbf{Implementation vs.\ production.} We test the open-source MarkLLM implementation of SynthID, not Google's proprietary production system. We cannot speak to how the production version would perform. The FRS is also one possible operationalization of forensic readiness---alternative designs might score differently.

\textbf{Framework validation breadth.} The FRS criteria and gates draw on Daubert and NIST, but we have not yet conducted inter-rater reliability testing or a systematic study of how different gate thresholds would affect outcomes. Those are next steps for maturing the framework.

\section{Conclusion}

All three watermarking methods we tested---KGW, Unigram, and the MarkLLM SynthID configuration---fall short of forensic evidence admissibility standards. Meaning-preserving paraphrase eliminates watermark detection in 100\% of initially-detected texts for KGW and Unigram, and 98.3\% for SynthID, while also revealing a 5.4\% false positive rate on clean text for SynthID. All three methods fail at least two of five Daubert factors. The FRS framework we propose gives courts and policymakers a structured tool for evaluating watermark evidence---while also revealing, through the Unigram boundary case, that point-based scoring has inherent limits.

The gap between what ``technical robustness'' means in an ML paper and what ``forensic admissibility'' means in a courtroom has to be closed before this evidence shows up at scale. Our framework is a first step. The alternative---letting unvalidated evidence into legal proceedings without anyone having checked it against the standards courts actually use---risks repeating the failures documented by the NAS \cite{nas2009} and PCAST \cite{pcast2016} reports. Those failures were measured in wrongful convictions.


\section{Ethical Considerations}

Everything in this paper evaluates publicly available systems. We use open-source models (Qwen2.5, Gemma-2) and a public toolkit (MarkLLM). No human subjects are involved. The attack we test---paraphrasing---is something anyone with internet access can do. We are not introducing new capabilities; we are documenting the forensic consequences of capabilities that already exist, which we believe serves the public interest.

\section{Researcher Positionality}

Our background is in digital forensics and adversarial security, not ML optimization. That shaped how we approached this work: we evaluated watermarks against the standards courts actually use to decide whether evidence is admissible, rather than measuring aggregate detection accuracy the way most ML robustness studies do. We think that framing is important because policymakers are already writing watermarking into law, and someone needs to ask whether the evidence holds up in the room where it will actually be used.

That said, this lens has blind spots. Watermarks may serve purposes beyond courtroom evidence — content provenance tracking, platform-level filtering, or internal audit — where the forensic bar we apply is not the right yardstick. Our evaluation does not speak to those use cases, and readers should keep that scope in mind.

\section{Adverse Impact Statement}

Showing that watermarks are fragile could undermine trust in watermarking prematurely, or be used to argue against mandates. But we think the greater risk is the alternative: deploying unreliable evidence technology in courts without anyone having validated it first. History supports this view. Delayed validation of forensic methods---bite mark analysis, hair microscopy, algorithmic risk scores like COMPAS \cite{loomis2016}---has led to documented wrongful convictions. Proactive validation, even when the results are negative, is the responsible path.


\section*{Acknowledgments}

We thank Alexander Nemecek for reviewing an earlier version of this work and for helpful discussion on watermark governance.

\bibliography{references}

\end{document}